\def\la{\hbox{{\lower -2.5pt\hbox{$<$}}\hskip -8pt\raise
-2.5pt\hbox{$\sim$}}}
\def\ga{\hbox{{\lower -2.5pt\hbox{$>$}}\hskip -8pt\raise
-2.5pt\hbox{$\sim$}}}
\def\ltsima{$\; \buildrel < \over \sim \;$}
\def\simlt{\lower.5ex\hbox{\ltsima}}
\def\gtsima{$\; \buildrel > \over \sim \;$}
\def\simgt{\lower.5ex\hbox{\gtsima}}

\documentstyle[epsfig]{elsart}

\begin{document}
\begin{frontmatter}
\title{The Small Scale Anisotropies, the Spectrum and the Sources of 
Ultra High Energy Cosmic Rays}

\author[inaf,infnfi]{Pasquale Blasi\thanksref{corr1}} 
\author[genova,infnge]{Daniel De Marco\thanksref{corr2}} 

\address[inaf]{INAF/Osservatorio Astrofisico di Arcetri,\\
Largo E. Fermi, 5 - 50125 Firenze, ITALY}
\address[infnfi]{INFN/Sezione di Firenze, ITALY}
\address[genova]{Universit\`a degli Studi di Genova,\\
Via Dodecaneso, 33 - 16146 Genova, ITALY}
\address[infnge]{INFN/Sezione di Genova, ITALY}

\thanks[corr1]{E-mail: blasi@arcetri.astro.it}
\thanks[corr2]{E-mail: ddm@ge.infn.it}

\begin{abstract} 
We calculate the number density and luminosity of the sources of ultra high 
energy cosmic rays (UHECRs), using the information about the small scale 
anisotropies and the observed spectra. We find that the number of doublets 
and triplets observed by AGASA can be best reproduced for a source density 
of $\sim 10^{-5}~\rm Mpc^{-3}$, with large uncertainties. The spectrum of 
UHECRs implies an energy input of $\sim 6\times 10^{44} \rm erg~yr^{-1}~
Mpc^{-3}$ above $10^{19}$ eV and an injection spectrum $\propto E^{-2.6}$. 
A flatter injection spectrum, $E^{-2.4}$, can be adopted if the sources have 
luminosity evolution $\propto (1+z)^4$.
The combination of these two pieces of information suggests that the single 
sources should on average have a cosmic ray luminosity above $10^{19}$ eV 
of $L_{source}\approx 2\times 10^{42} \rm erg~s^{-1}$, weakly dependent 
upon the injection spectrum. Unfortunately, with the limited statistics of 
events available at present, there are approximately one-two orders of 
magnitude uncertainty in the source density provided above. We make 
predictions on the expected performances of the Auger and EUSO experiments, 
with particular attention for the expected improvements in our
understanding of the nature of the sources of UHECRs. We find that 
a critical experimental exposure $\Sigma_c$ exists, such that experiments 
with exposure larger than $\Sigma_c$ can detect at least one event from each 
source at energies above $10^{20}$ eV. This represents a unique opportunity 
to directly count and identify the sources of UHECRs.
\end{abstract}
\end{frontmatter}

\section{Introduction}

Despite the efforts in building several large experiments for the
detection of particles with the highest energies in the cosmic ray
spectrum, the nature of the sources of these particles and the
acceleration processes at work are still unknown. No trivial counterpart
has been identified by any of the current experiments: there is no
significant association with any large scale local structure nor with
single sources. However, caution should be adopted in looking for the
sources and in claiming their absence: particles with energy around
$4\times 10^{19}$ eV have a loss length of $\sim 1000$ Mpc, 
therefore many sources lie in the $\sim 2-3$ degrees of angular
resolution of current experiments. It follows that the identification
of one of them as responsible for the production of a single particle
can turn out to be very difficult. In fact, for particles with this
energy, one could say that there are too many possible sources that may
potentially contribute, rather than the opposite. On the other hand, the
need for statistically significant findings forces current experiments
to work in this energy region. The events with energy above a few
$10^{20}$ eV, where the loss length for photopion production is only
$\sim 20-50$ Mpc are only a handfull, but at least in principle it
should be much simpler to find their sources. For instance, if we
assumed that sources located within ordinary galaxies could accelerate
particles to energy in excess of $10^{20}$ eV, one could easily estimate
that in an angle of $2$ degrees, at most $\sim 1$ galaxy should be
within the near $20$ Mpc from us. On the other hand, for particles with
a pathlength $1000$ Mpc, in the same angular bin there would be $\sim
4\times 10^4$ galaxies.

It has been recently proposed that a statistically significant
correlation may exist between the arrival directions of UHECRs with
energy above $2\times 10^{19}$ eV and the spatial location of BL Lac objects,
with redshifts larger than 0.1 \cite{tkachev1,tkachev2}. Since at these 
energies the loss length is comparable with the size of the universe, no 
New Physics needs to be invoked to explain the correlation. On the other
hand, since no BL Lac is known to be located close to the Earth, the 
spectrum of UHECRs expected from these sources would have a very pronounced
GZK cutoff. 

At the highest energies even the spectrum of UHECRs is measured rather poorly.
The low expected number of events above $10^{20}$ eV does not allow a clear
identification of the so-called GZK feature, caused by the sharp drop in the
loss length of UHECRs at the onset of photopion production as the dominant
channel of energy losses. In \cite{daniel1} it was shown that the two largest 
experiments, namely AGASA and HiRes \cite{HIRES1,HIRES2}, cannot measure 
the presence or lack of the GZK feature at a statistical level better than 
$\sim 2 \sigma$. In the present paper, all the numbers quoted as expected
numbers of events refer to the propagation of cosmic ray protons from 
astrophysical sources, therefore a GZK feature is present in all the spectra. 
If future experiments will show the absence of the GZK feature in the spectrum 
of UHECRs, then many more events will be detected than predicted in this paper.

The potential for discovery of the sources has recently improved after 
the identification of a few doublets and triplets of events clustered on 
angular scales comparable with the experimental angular resolution of 
AGASA. A recent analysis of the combined results of most UHECR experiments 
\cite{uchihori} revealed 8 doublets and two triplets on a total of 92 events 
above $4\times 10^{19}$ eV (47 of which are from AGASA). The recent HiRes
data do not show evidence for significant small angle clustering, but this
might well be the consequence of the smaller statistics of events and of 
the energy dependent acceptance of this experiment: one should remember 
that in order to reconstruct the spectrum of cosmic rays, namely to account 
for the unobserved events, a substantial correction for this energy dependence 
need to be carried out in HiRes (the acceptance is instead a flat function
of energy for AGASA at energies above $10^{19}$ eV). This correction however 
does not give any information on the spatial distribution of the {\it missed} 
events.

If the appearance of these multiplets in the data will be confirmed by future 
experiments as not just the result of a statistical fluctuation or focusing in 
the galactic magnetic field \cite{harari}, then the only way to explain their 
appearance is by assuming that the sources of UHECRs are in fact point sources.
This would represent the first true indication 
in favor of astrophysical sources of UHECRs, since the clustering of events 
in most top-down scenarios for UHECRs seems unlikely. A clear identification
of the GZK feature in the spectrum of UHECRs would make the evidence in favor
of astrophysical sources even stronger.

Previous attempts to estimate the number of sources of UHECRs in our cosmic
neighborhood from the small scale anisotropies found by AGASA have been carried
out, adopting both semi-analytical and numerical approaches. An analytical
tool to evaluate the chance coincidence probability for arbitrary statistics
of events was proposed in \cite{tom}. A rigorous analysis of the clusters
of events and of their energy dependence was given in \cite{will}. 
In \cite{dubovsky} the authors use an analytical method to estimate the 
density of the sources of UHECRs restricting their attention to the 14 events 
with energy above $10^{20}$ eV with one doublet. They obtain a rather 
uncertain estimate centered around $6\times 10^{-3}~\rm Mpc^{-3}$. In 
\cite{fodor} the energy losses are introduced through a function, derived 
numerically, that provides the probability of arrival of a particle from a 
source at a given distance.
Again, only events above $10^{20}$ eV are considered, therefore the analysis
is based upon one doublet of events out of 14 events. This causes extremely
large uncertainties in the estimate of the source density, found to be
$180^{+2730}_{-165} \times 10^{-3}~\rm Mpc^{-3}$. No account of the 
statistical errors in the energy determination nor of the declination 
dependence of the acceptance of the experimental apparata is included in 
all these investigations. In the present paper we address the issue of 
calculating the number density and luminosity of the sources of UHECRs
with a full numerical simulation of the propagation, and we take into
account the statistical errors in the energy determination and the
declination dependence of the experiments involved. We carry out our analysis
for cosmic rays detected by AGASA, with energy above $4\times 10^{19}$ eV, 
where the statistics of events is more generous. We also apply the same
analysis to mock catalogs of events from Auger \cite{auger} and EUSO 
\cite{EUSO}, with the statistics of events expected for each. Different 
tools are proposed to determine with some accuracy the density of the 
sources of UHECRs.
We will present our results with magnetic fields in a forthcoming paper,
extending recent interesting findings reported in \cite{sato1,sato2}
to the case of time dependent sources of UHECRs. The propagation of
UHECRs in magnetic fields was also investigated in \cite{Stanevmag},
where the important concept of magnetic horizon was introduced, and
in \cite{isola,ensslin} where special attention was devoted to the
local neighborhood, although the contribution of distant sources was
not considered.

The paper is organized as follows: in \S \ref{sec:propsim} we provide a 
description of the simulations used to propagate particles
from point sources distributed throughout the universe.
In \S \ref{sec:sources} we use observational data and simulations to constrain
the number of sources in the near universe and their luminosity. We also 
discuss at length several tools that may be used with data from Auger
and EUSO to infer the nature of the sources of UHECRs. In \S \ref{sec:single}
we discuss the possibility of using future EUSO data to measure the 
spectrum of a single source of cosmic rays with energy above 
$5\times 10^{19}$ eV. In \S 
\ref{sec:magnetic} we present a critical view of the role of magnetic 
fields for the propagation of UHECRs, in order to define the limits of 
the results reported in this paper. We conclude in \S \ref{sec:conclude}.

\section{The Montecarlo simulation for cosmic ray propagation}
\label{sec:propsim}

The propagation of UHECRs is simulated here using the Montecarlo 
simulation described in a previous paper \cite{daniel1}. We refer 
the reader to that paper for more details, while in this section 
we provide the basic information.

We assume that ultra-high energy cosmic rays are protons injected with a
power-law spectrum  in extragalactic sources. The injection spectrum is
taken to be of the form
\begin{equation}
F(E) d E = \alpha E^{-\gamma} \exp(-E/E_{\rm max}) d E
\label{eq:inj}
\end{equation}
where $\gamma$ is the spectral index, $\alpha$ is a normalization constant, 
and $E_{\rm max}$ is the maximum energy at the source. 

We simulate the propagation of protons from source to observer by including the
photo-pion production, pair production, and adiabatic energy losses due to
the expansion of the universe \cite{Blanton,daniel1}, treating photo-pion 
production as a discrete energy loss process.

In each step of the simulation, we calculate the pair production losses
using the continuous energy loss approximation given the small inelasticity in
pair production ($2 m_{\rm e}/m_{\rm p}\simeq10^{-3}$).  For the rate of
energy loss due to pair production at redshift $z=0$, $\beta_{\rm pp}(E,z=0)$,
we use the results from \cite{Blumenthal:nn,czs}, while at redshift $z>0$,
\begin{equation}
\beta_{\rm pp}(E,z)=(1+z)^3 \beta_{\rm pp}((1+z)E, z=0)\,.
\end{equation}
Similarly, the rate of adiabatic energy losses due to redshift is
calculated in each step  using
\begin{equation}
\beta_{\rm rsh}(E,z)=H_0 \left[\Omega_M (1+z)^3 +
\Omega_\Lambda\right]^{1/2}~,
\end{equation}
with $H_0=75 ~ {\rm km~ s^{-1} Mpc}^{-1}$.

The photo-pion production is simulated by calculating first the average 
number of photons able to interact via photo-pion production
through the expression:
\begin{equation}
\langle N_{\rm ph}(E,\Delta s) \rangle=\frac{\Delta s}{l(E, z)},
\end{equation}
where $l(E,z)$ is the interaction length for photo-pion production of a
proton with energy $E$ at redshift $z$ and $\Delta s$ is a step size,
chosen to be much smaller than the interaction length (typically we
choose $\Delta s=100 ~{\rm kpc}/(1+z)^3$).

\begin{figure}
\begin{center}
\includegraphics[width=0.7\textwidth]{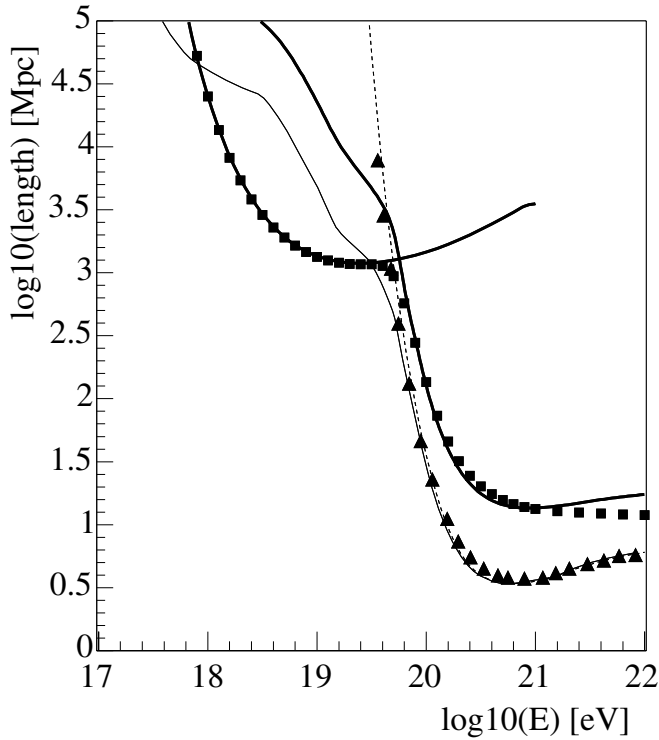}
\caption{Interaction length for photopion production as calculated in this
paper (dashed line) compared to the interaction length of \cite{BS00}
(solid thin line) and of \cite{Stanev} (triangles). The thick solid lines
are the loss lengths for photopion production (on the right) and of pair
production (on the left) as obtained in \cite{BS00}. The squares are the loss
lengths for photopion production and proton pair production as calculated
in \cite{bgg2002}.}
\label{fig:il}
\end{center}
\end{figure}

In Fig. \ref{fig:il} we plot the interaction length for photopion
production used in \cite{BS00} (solid thin line), and in \cite{Stanev}
(triangles). The dashed line is the result of our calculations (see
below), which is in perfect agreement with the results of
\cite{BS00,Stanev}. The apparent discrepancy at energies below
$10^{19.5}$ eV with the prediction of Ref. \cite{BS00} is only due to
the fact that we consider only microwave photons as background, while in
\cite{BS00} the infrared background was also considered. For our
purposes, this difference is irrelevant as can be seen from  the loss
lengths plotted in Fig. \ref{fig:il}. The rightmost thick solid line is
the loss length for photopion production \cite{BS00}, while the other
thick solid line is the loss length for pair production. For comparison,
in Fig. \ref{fig:il} we also plot the loss length as calculated in 
\cite{bgg2002} (thick squares).
In the present calculations, we do not use the loss length of photopion
production which is related to the interaction length through an angle
averaged inelasticity. In our simulations we evaluate the inelasticity
for each  single proton-photon scattering using the kinematics, rather
than adopting an angle averaged value.

The interaction length is calculated as described in detail in \cite{daniel1}.
Once the interaction length is known, we then sample a Poisson
distribution with mean $\langle N_{\rm ph}(E,\Delta s) \rangle$, to determine
the  actual number of photons encountered during the step $\Delta s$. When a
photo-pion interaction occurs, the energy $\epsilon$ of the photon is extracted
from the Planck distribution, $n_{ph}(\epsilon,T(z))$, with temperature
$T(z)=T_0 (1+z)$, where $T_0=2.728$ K is the temperature
of the cosmic microwave background at present. Since the microwave photons
are isotropically distributed, the interaction angle, $\theta$, between the
proton and the photon is sampled randomly from a distribution which is flat
in $\mu={\rm cos}\theta$. Clearly only the values of $\epsilon$ and
$\theta$ that generate a center of mass energy above the threshold for
pion production are considered. The energy of the proton in the final
state is calculated at each interaction from kinematics.
The simulation is carried out until the statistics of events
detected above some energy reproduces the experimental numbers.
In this way we have a direct handle on the fluctuations that can be
expected in the observed flux due to the stochastic nature of photo-pion
production and to cosmic variance.
In the presence of a homogeneous distribution of point sources the 
simulation proceeds in the following way: we fix the density of sources
that we wish to use for the simulation and we generate at random the 
positions of the point sources in all the universe. For our calculations
we adopt a flat $\Lambda$-dominated universe with $\Omega_\Lambda=0.7$ 
and $\Omega_m=0.3$. In a Euclidean universe, the flux from a source would
scale as $r^{-2}$ where $r$ is the distance between the source and the
observer. On the other hand, the number of sources between $r$ and $r+dr$ would
scale as $r^2$, so that the probability that a given event has been
generated by a source at distance $r$ is independent of $r$: sources
at different distances have the same probability of generating any
given event. In a flat universe with a cosmological constant, this is
still true provided the distance $r$ is taken to be
\begin{equation}
r = c \int_{t_g}^{t_0} \frac{dt}{R(t)},
\end{equation}
where $t_g$ is the age of the universe when the event was generated, $t_0$
is the present age of the universe, and $R(t)$ is the scale factor.

Once a source distance has been selected at random, one of the sources
at that distance is picked at random and a particle energy is assigned 
to the event from a distribution that reflects the injection spectrum, 
chosen as in Eq. (\ref{eq:inj}). This particle is then propagated to the 
observer and its energy and direction of arrival recorded. Due to the
statistical error in energy determination, the recorded energy of the event
is generated from a gaussian centered around the arrival energy with a
spread defined by the statistical error. A similar procedure is adopted to 
account for the uncertainty in the direction of arrival, determined by
the angular resolution of the instrument. This procedure
is repeated until the number of events above a threshold energy, $E_{th}$
is reproduced. With this procedure we can assess the significance of
results from present experiments with limited statistics of events.

The successful results of the comparison of our findings with 
the results of our simulations with the analytical results 
\cite{beregrig,berebook,bereAGN,bgg2002} valid in the low energy 
regime (namely below $4\times 10^{19}$ eV) have already been discussed 
in \cite{daniel1}.

\section{How many sources?}\label{sec:sources}

In this section we illustrate the results of our calculations for the
simulated number of doublets and triplets obtained using the events
statistics of the AGASA experiment with the declination dependence of the
acceptance given in \cite{takeda}. While the current statistics of
doublets and triplets is richer, the exact arrival directions are not
publically available, therefore we use here the numbers of multiplets
reported in \cite{agasa1}, for which clear information about the directions 
of arrival is provided. 
We consider several values for the source density in the local universe and 
two cases of source evolution with redshift, namely no evolution and evolution 
as $(1+z)^m$ with $m=4$. The location of the sources is generated at random 
with the chosen value of the spatial density. UHECRs are then generated from
these sources and propagated to the Earth as explained in \S \ref{sec:propsim}.
The angle dependence of the acceptance of the experiment is also taken 
into account. 
After generating the statistics of events observed by the experiment
above some energy threshold, we calculate the number of n-plets. This
procedure is repeated for 100 realizations of the source distribution
in order to calculate the average expected numbers and the corresponding
uncertainties.

A first information about the small and possibly large scale anisotropy 
of the AGASA data can be extracted from the two point correlation function of
the observed data, defined as \cite{sato2,isola}: 
\begin{equation}
N(\theta) = \frac{1}{S(\theta)}\sum_{i>j} R_{i j}(\theta),
\end{equation} 
where $S(\theta)=2\pi |\cos(\theta)-\cos(\theta+\Delta \theta)|$ is the area 
of the angular bin between $\theta$ and $\theta+\Delta \theta$ and 
$R_{i j}(\theta)=0,1$ counts the events in the same bin.

\begin{figure}
\begin{center}
\includegraphics[width=1.1\textwidth]{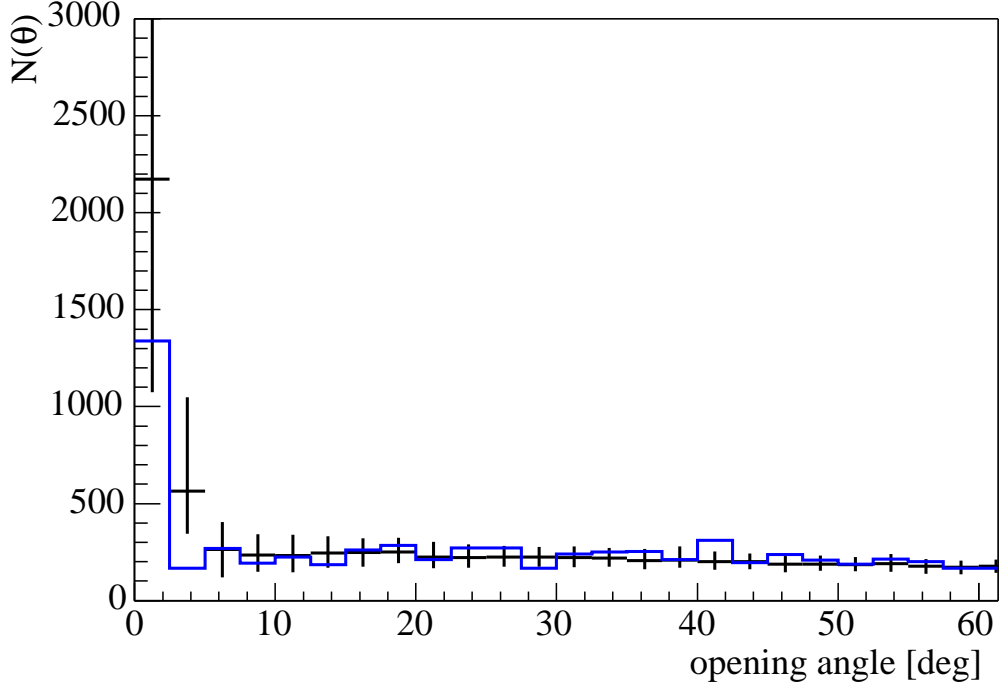}
\caption{Two Point correlation function of the AGASA data (histogram) compared
with the prediction of our simulations (points with error bars) averaged
over 100 realizations.}
\label{fig:agasadata}
\end{center}
\end{figure}

The two point correlation function of the AGASA data is plotted as a histogram
in Fig. \ref{fig:agasadata} for angles between 2 and 60 degrees. In the 
same plot the points with error bars are the result of our simulations 
with the AGASA statistics of events, averaged over 100 realizations of 
the source distribution with a source density of $10^{-5}~\rm Mpc^{-3}$
(see discussion below).
The presence of clusters of events reflects into the appearance of 
the large peak at the angular resolution of AGASA, while no additional 
feature is present in the two point correlation function.

\begin{figure}
\begin{center}
\includegraphics[width=1.1\textwidth]{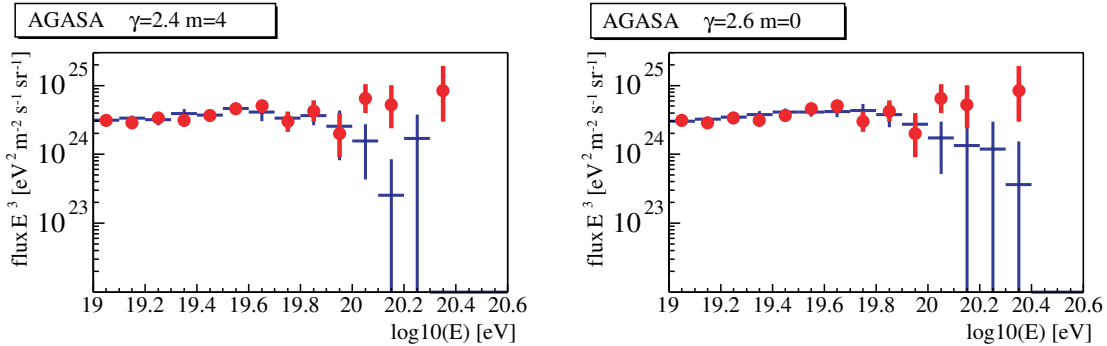}
\caption{Observed spectrum (thick dots with error bars) compared
with the results of our calculations, using an injection spectrum $E^{-2.4}$
and redshift evolution of the sources $(1+z)^4$ (left panel) or an 
injection spectrum $E^{-2.6}$ with no evolution (right panel).}
\label{fig:AGASAspec}
\end{center}
\end{figure}

We can now use our simulations to calculate the numbers of multiplets
with multiplicities $n=2,~3,...$ for the same energy threshold as in 
the AGASA experiment, namely $4\times 10^{19}$ eV. The calculations 
are carried out for an opening angle of 3 degrees and an angular 
resolution of the AGASA experiment of 2.5 degrees. The observed spectrum 
of UHECRs above $10^{19}$ eV can be fitted equally well with an injection 
spectrum $E^{-2.6}$ and no redshift evolution, or with an injection 
spectrum $E^{-2.4}$ and redshift evolution $\propto (1+z)^4$. We carry
out our simulations for both cases. The diffuse spectra that we obtain are
plotted in Fig. \ref{fig:AGASAspec} for the two cases, compared with
AGASA data (thick dots).

\begin{figure}
\begin{center}
\includegraphics[width=1.1\textwidth]{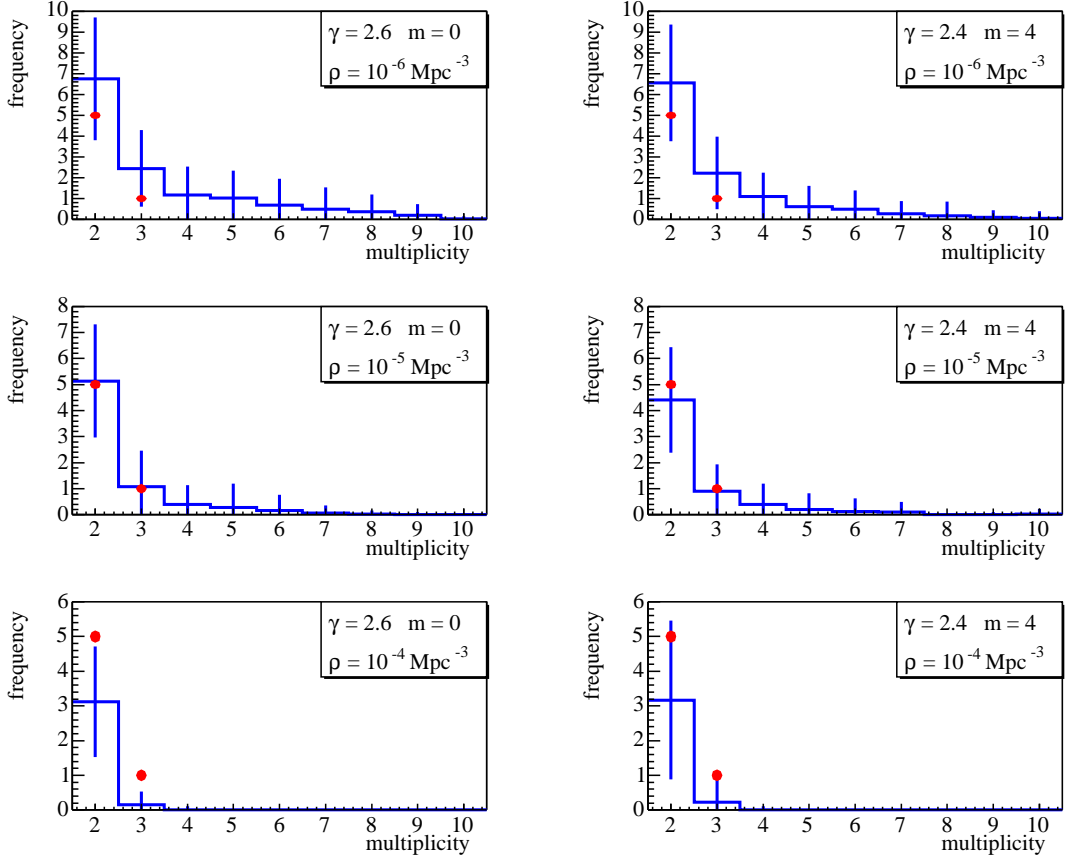}
\caption{Statistics of multiplets with the AGASA statistics as obtained
in our simulations with the source densities indicated in the plots. The 
left column refers to the case without redshift evolution, while the right
column refers to the case with evolution $(1+z)^4$ in the injection spectra.
The dots are the AGASA findings.}
\label{fig:AGASAmulti}
\end{center}
\end{figure}

The calculations of the small scale anisotropies are carried out for 
three source densities $\rho=10^{-6}$, $\rho=10^{-5}$ and 
$\rho=10^{-4}~\rm Mpc^{-3}$. 
Our results are illustrated in Fig. \ref{fig:AGASAmulti}: the plots on
the left side are obtained with injection spectrum $E^{-2.6}$ and no
evolution ($m=0$), while the plots on the right are for injection
spectrum $E^{-2.4}$ and $m=4$ (strong evolution). These two cases are
likely to bracket the region of plausible evolution of the sources. 
As we show below however, the evolution in the source luminosity or
density does not affect in any appreciable way the small angle clustering 
of UHECRs. The value of the source density adopted in our calculations is 
indicated in the plots. 

The multiplets seen by the AGASA experiment consist of 5 doublets and
1 triplet, indicated in Fig. \ref{fig:AGASAmulti} with two thick dots
at multiplicities 2 and 3. 
For the case of low source density, $\rho=10^{-6}~\rm Mpc^{-3}$, 
there are few sources from which an event can be generated. As a consequence,
the multiplets are frequent. There is a finite probability in
this case to have multiplets with multiplicity up to 6, which are not
observed. The case with source density $\rho=10^{-5}~\rm Mpc^{-3}$
provides a better description of the data, namely the numbers of doublets 
and triplets appear to be closer to the observed numbers and the higher
multiplicity clusters are less frequent. For the higher 
density case, an increasing number of sources can contribute events, 
therefore doublets and triplets, and clearly the higher multiplicity 
multiplets as well, become more sparse.

We can conclude that for continuous (namely non-bursting) sources of 
UHECRs, the expected source density should be 
$\rho\approx 10^{-5}~\rm Mpc^{-3}$,
with an uncertainty of more than one order of magnitude (a few - $400$
sources within 100 Mpc from the Earth).
The number is approximately the same for the two cases of injection
spectrum and evolution. This could be expected because the sources that 
contribute a flux at energies above $4\times 10^{19}$ eV are those at 
redshift $z\leq 0.2-0.3$, where the evolution is still only marginally 
relevant. 

It is worth stressing once more that the two point correlation function 
of the realizations that provide the doublets and triplets with the 
AGASA statistics, as shown in Fig. \ref{fig:agasadata} do not 
show evidence for features other than the peak at small angles.
In the following we consider the situation that we expect to take
place for Auger and EUSO.

\par\noindent
{\it The case of the Auger Observatory}

We consider here the Auger observatory in the south emisphere, with 
total acceptance of 7000 $\rm km^2~sr$, and declination dependence as
given in \cite{sommers}. This number has to be compared with the AGASA
acceptance of $160~\rm km^2~sr$. AGASA has detected 886 events above 
$10^{19}$ eV, with an exposure of $1645~\rm km^2~sr~yr$. In 5 years 
of operation of Auger, this would correspond to $\sim 1.9\times 10^4$
events above $10^{19}$ eV and $1.5\times 10^3$ events above $4\times 10^{19}$
eV. The expected number of events  above $10^{20}$ eV is more dependent 
on details of the cosmic ray propagation. For an injection spectrum 
$E^{-2.6}$ and with no luminosity evolution of the sources, we predict 
$\sim 60-70$ events with energy above $10^{20}$ eV. About $10\%$ of the 
events are expected to be detected by both the ground array and the 
fluorescence telescopes.

\begin{figure}
\begin{center}
\includegraphics[width=1.1\textwidth]{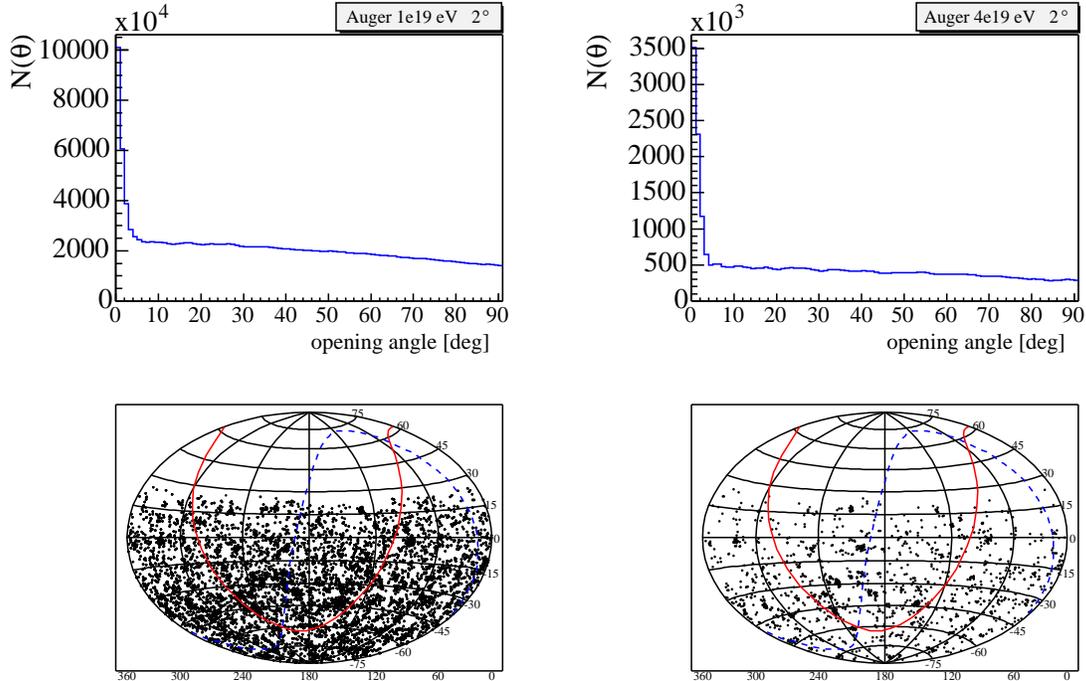}
\caption{Upper plots: two point correlation function for the Auger expected 
statistics above $10^{19}$ eV (on the left) and above $4\times 10^{19}$ eV
(on the right). The lower plots are the corresponding sky maps for one
realization of the source distribution.}
\label{fig:Auger}
\end{center}
\end{figure}

In Fig. \ref{fig:Auger} we plot the two point correlation function obtained
for Auger above $10^{19}$ eV (left upper plot) and above $4\times 10^{19}$ 
eV (right upper plot) with 2 degrees angular resolution, and using
a source density of $10^{-5} \rm Mpc^{-3}$. The lower plots represent 
one realization of the sky in the same energy region as in the corresponding
upper plot (the realization provides the numbers of doublets and triplets
currently seen by AGASA. The solid line identifies the supergalactic plane
while the dashed line locates the galactic plane). Again, no feature is present
in the two point correlation function with the exception of the peak at small 
angles.

\begin{figure}
\begin{center}
\includegraphics[width=0.8\textwidth]{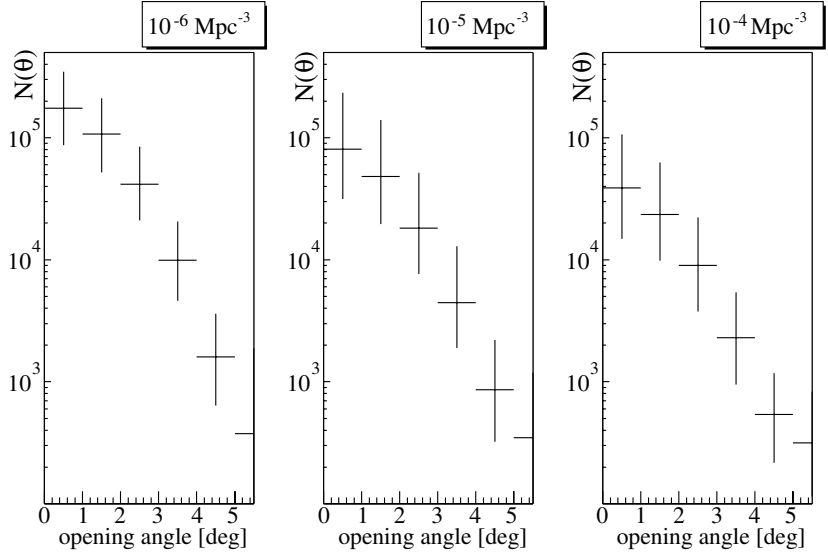}
\caption{Two Point correlation function for the Auger simulated data above 
$10^{20}$ eV for source density $10^{-6}$, $10^{-5}$ and 
$10^{-4}~\rm Mpc^{-3}$.}
\label{fig:Auger2pcf}
\end{center}
\end{figure}

Although these plots refer to a specific realization, it can be shown
that the correlation function averaged over many realizations has a similar 
shape and the same peak at small angles, but with large error bars. The reason 
for these uncertainties is that with the expected number of events above 
$4\times 10^{19}$ eV ($\sim 1500$), the angular distance between two events 
in the sky is $\sim 2$ degrees, comparable with the point spread function of 
the experiment.
In these circumstances the accidental clusters of events, namely those that 
are not associated with a point source, are very frequent. In
this respect the correlation function of the events above $4\times 10^{19}$ 
eV might not be the best tool to count the sources of UHECRs. 
We make the attempt to apply the same method restricted to events above
$10^{20}$ eV. Our results for the two point correlation function are plotted
in Fig. \ref{fig:Auger2pcf} for three values of the source density, 
namely $10^{-6}$, $10^{-5}$ and $10^{-4}~\rm Mpc^{-3}$. The error bars
are obtained after averaging the propagation over 100 realizations of
the source distribution. 
On the basis of the size of the error bars, it appears that it should be
reasonably simple with Auger data above $10^{20}$ eV to infer the 
source density within at least one order of magnitude, which is a clear
improvement on what is possible to achieve with AGASA data.

\begin{figure}
\begin{center}
\includegraphics[width=0.8\textwidth]{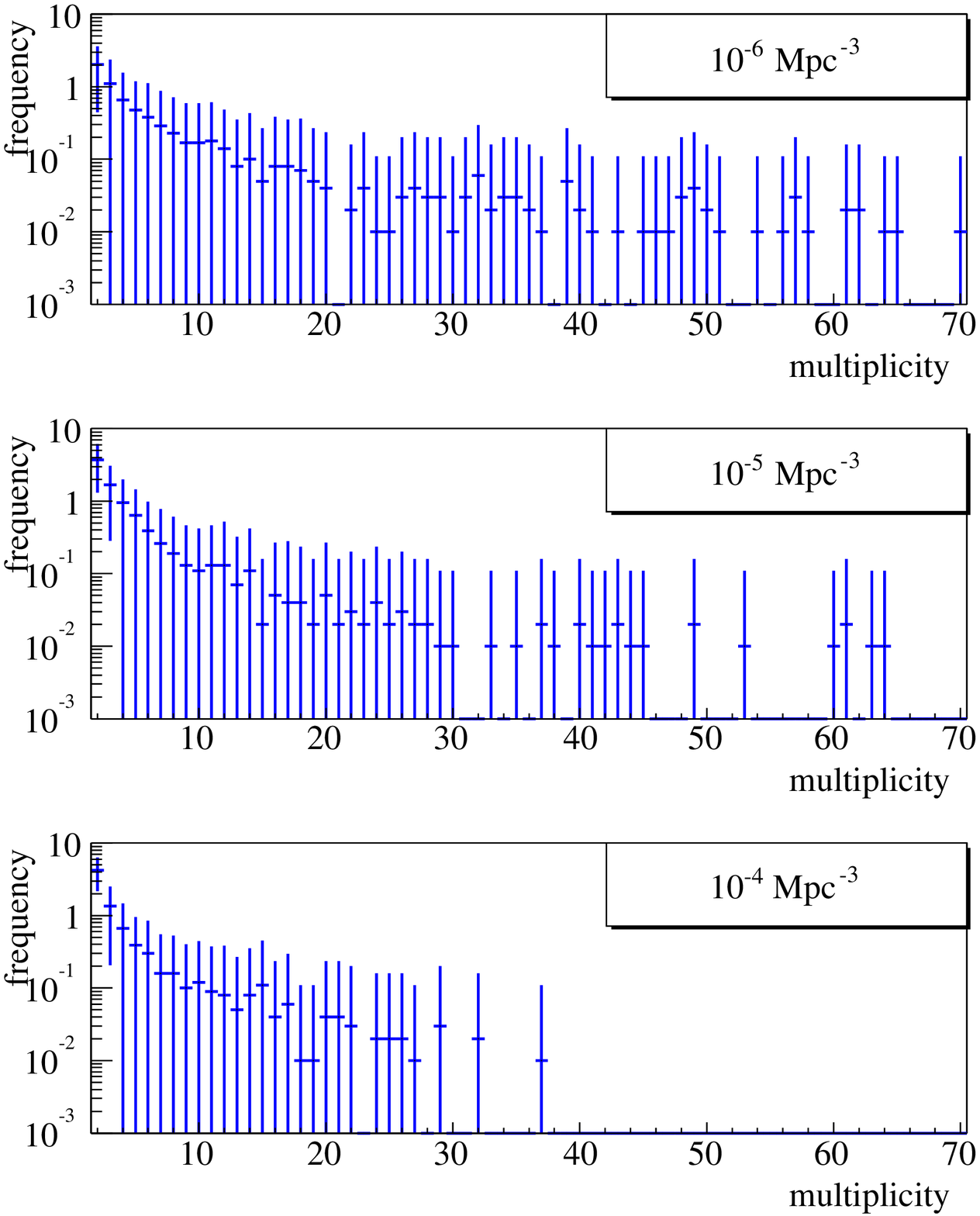}
\caption{Multiplets of events in the Auger simulated data above 
$10^{20}$ eV for source density $10^{-6}$, $10^{-5}$ and 
$10^{-4}~\rm Mpc^{-3}$.} 
\label{fig:Augermulti}
\end{center}
\end{figure}
In Fig. \ref{fig:Augermulti} we plot the average number of clusters of events
with different multiplicity and related error bars, as expected above 
$10^{20}$ eV in Auger, for the three values of the density of sources.
The error bar in the number of doublets appears to be small enough to allow
one to determine the number of sources by following the same procedure 
illustrated above for the case of AGASA. Note that here the multiplets are
counted in a {\it non compact} manner, namely a set of $N$ events qualifies
as a N-plet even if not all the pairs of events within it have an angular 
separation less than some value which is fixed in the analysis. This
creates some ambiguity in the definition of a cluster, but more rigorous
definitions can be easily found. We adopted this definition in order to 
avoid to check the angular distances of all the pairs of events in a 
multiplet with very high multiplicity, since this can easily become an 
unfeasible task for large values of $N$.

It is worth noticing that the
tendency here is the opposite with respect to that obtained for the 
AGASA statistics: the number of doublets increases with increasing source
density. At first this may seem counterintuitive, because with a larger number
of sources to generate an event from, it is likely that each time a 
different source is picked. On the other hand, when the number of events
($\sim 60-70$ for Auger above $10^{20}$ eV) becomes comparable with 
the number of sources that can contribute the events, the
clustering becomes more probable and higher multiplicity clusters appear. 
This reflects in the fact that the ratio of the number of events that come 
in singlets to the number of clustered events increases when the source 
density increases.

\par\noindent
{\it The case of EUSO}

The EUSO acceptance is known within a factor of $\sim 2$ and amounts
to $35000-70000~\rm km^2~sr$. The observation time is supposed to be 
3 years, therefore the total exposure is expected to be of 
$\sim (1-2)\times 10^5 \rm km^2~sr~yr$. If the GZK feature is in 
fact present in the cosmic ray data, the expected number of events
above $10^{20}$ eV in EUSO is $\sim 180-360$ if the maximum energy 
of the particles at the sources is large enough.

\begin{figure}
\begin{center}
\includegraphics[width=1.1\textwidth]{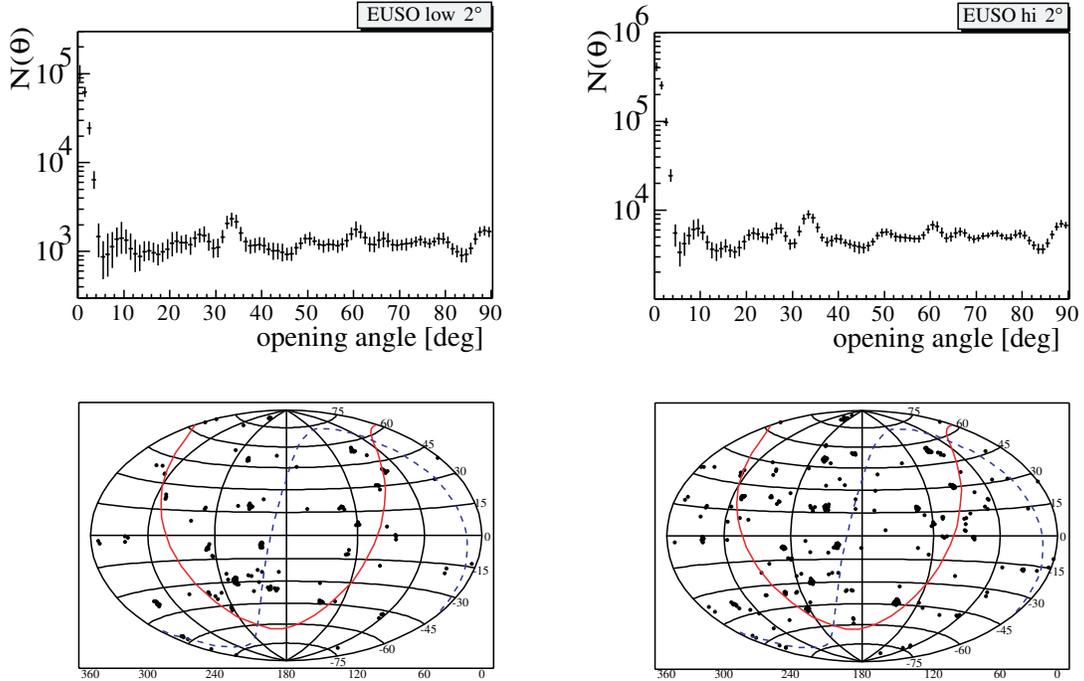}
\caption{Upper plots: two point correlation function for the EUSO expected 
statistics above $10^{20}$ eV for its baseline configuration (left)
and twice as much (right). The lower plots are the corresponding sky 
maps for one realization of the source distribution.} 
\label{fig:EUSO_4panels}
\end{center}
\end{figure}

In the following we consider two cases for the EUSO acceptance,
with average number of events above $10^{20}$ eV equal to $180$
(low acceptance case) and $360$ (high acceptance case). 
The correlation function for the low (left upper plot) and high acceptance 
(right upper plot) cases is plotted in Fig. \ref{fig:EUSO_4panels} for a 
density of sources $10^{-5} \rm Mpc^{-3}$. The corresponding sky is 
illustrated in the lower plots in  Fig. \ref{fig:EUSO_4panels}.

In the large angle region, where no feature was found for the AGASA and 
Auger statistics of events, the simulated data for the EUSO
experiment show evidence for some wiggles around the average value of
the two point correlation function at large angles, that is $\sim 
N_{ev}^2/8\pi$, where $N_{ev}$ is the number of events used for 
the analysis. 
The error bars are calculated by generating 100 realizations of 
the propagation of UHECRs, at fixed configuration of the sources
and at fixed number of events above some threshold. The size of the
error bars shows that these features should be visible in the two point
correlation function even after accounting for uncertainties in the
particle propagation. In other words, the peaks are not due to limited
statistics of events in the energy region of interest.
A similar plot for the Auger expected number of events above $10^{20}$
eV shows error bars which hide almost completely the wiggles in the 
correlation function.

\begin{figure}
\begin{center}
\includegraphics[width=0.8\textwidth]{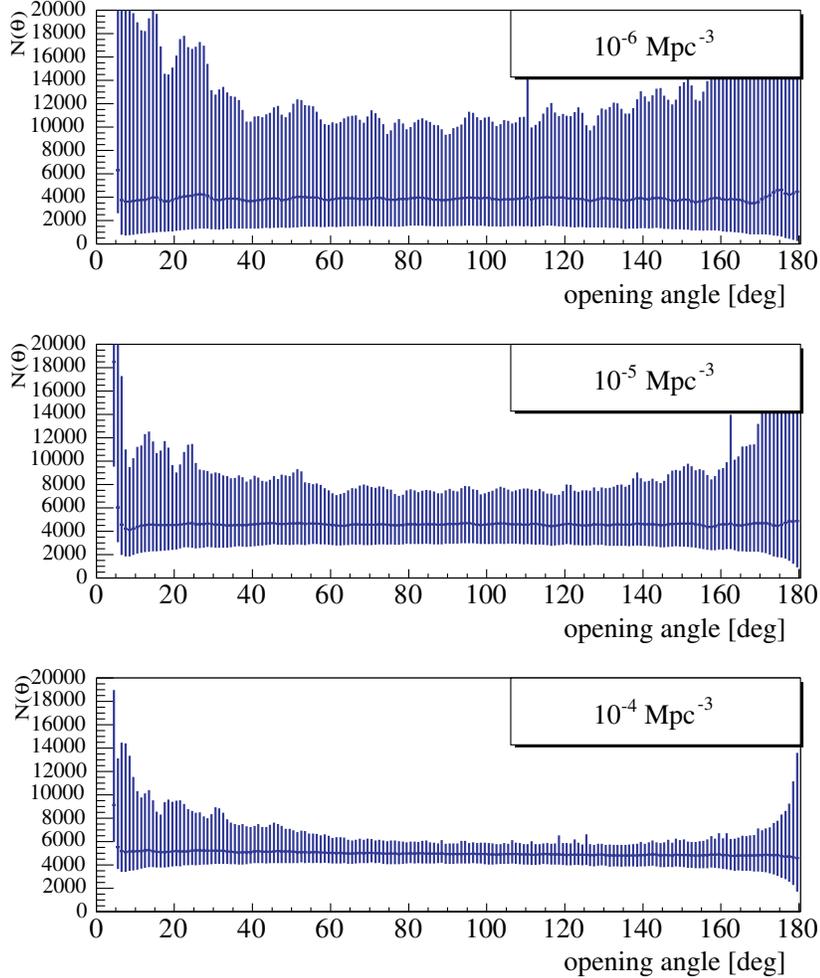}
\caption{Two point correlation function for the EUSO statistics of events
above $10^{20}$ eV, in the high acceptance case. The density of sources is
as indicated in the panels. The error bars are obtained by averaging the
correlation function on 2200 realizations of the source distribution.}
\label{fig:EUSOfluctua}
\end{center}
\end{figure}

\begin{figure}
\begin{center}
\includegraphics[width=0.8\textwidth]{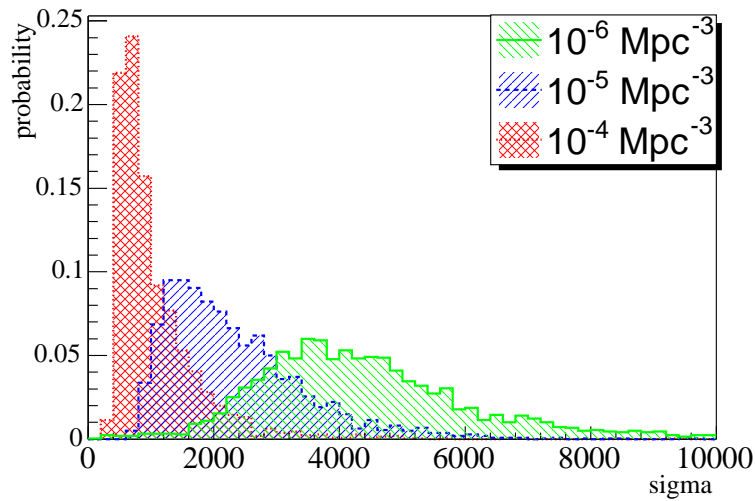}
\caption{Distribution of the variances around the mean of the two point
correlation function for the three values of the source density, as 
indicated. The histograms are obtained by averaging over 2200 realizations
of the source distribution.}
\label{fig:sigma}
\end{center}
\end{figure}

The wiggles are the consequence of random noise in the source 
distribution: this can be understood in terms of fluctuation of the
number of sources that may contribute events in the angular bin between
$\theta$ and $\theta+\Delta \theta$. According with this interpretation, one 
expects that the amplitude of these fluctuations gets larger for
lower source densities. Note that we are calculating the two point correlation
function of the events and not of the sources, therefore the information
on the sources is embedded in the two point correlation function and needs 
to be disentangled from the information on the events. On the other hand,
with the EUSO statistics of events the fluctuations due to the propagation
have been shown to be irrelevant, therefore the remaining fluctuations 
have to be due to the random distribution of the sources. The question 
we address here is whether we can use the amplitude of the fluctuations
around the average of the two point correlation function as an indicator of
the density of sources. In Fig. \ref{fig:EUSOfluctua} we plot the two point
correlation function for three values of the source density, namely
$\rho=10^{-6}$, $\rho=10^{-5}$ and $\rho=10^{-4}~\rm Mpc^{-3}$ averaged 
over 2200 realizations of the source distribution. 
One can clearly see that the scatter in the large angles part of the 
two point correlation fuction around its average value are much larger for 
the low density case than for the higher density cases, as expected. 
One can calculate for each realization of the sources the average 
variance of the two point correlation function around the expected average
at angles between 20 and 160 degrees and then plot the distribution of 
variances, for the three values of the source density. 
This graph is shown in Fig. \ref{fig:sigma} for 2200 realizations of the 
source distribution.
As an instance, the realization used to obtain Fig. \ref{fig:EUSO_4panels}
(source density $10^{-5}~\rm Mpc^{-3}$) had average variance around the
mean equal to 1200. This value is just on the right of the peak of the 
distribution of variances for a source density $10^{-4}~\rm Mpc^{-3}$ and
on the immediate left of the peak for source density $10^{-5}~\rm Mpc^{-3}$.
The probability to get a larger (smaller) variance in the two cases 
respectively is $28\%$ ($20\%$). The value of the variance in our specific 
realization is on the very tail of the distribution for 
$10^{-6}~\rm Mpc^{-3}$, with a probability of getting less of only $2\%$. 

The appearance of the wiggles in the two point correlation function is the
result of an interesting effect, peculiar of the ultra high energy
cosmic rays with energy above $10^{20}$ eV. The point-like nature of 
the sources arises in the two point 
correlation function when the number of events equals or exceeds the
number of sources that can contribute events in the energy region of
interest. For instance, for the AGASA data above $4\times 10^{19}$ eV
the number of events is a few tens but the sources that can contribute 
are all the sources within $\sim 1000$ Mpc. For the densities that we
found, the number of events is always much smaller than the number of
sources that generate them. In other words most sources did not contribute
any events at the detector. For Auger, we expect $\sim 1500$ events 
at energies above $4\times 10^{19}$ eV, while the sources that may 
contribute are $2\times 10^4$ for source density $10^{-5}~\rm Mpc^{-3}$
(Auger sees about half of the sky)
which is always larger than the number of events for the three source
densities that we restricted our attention to. Moreover, even from a
visual look at the sky as simulated for the Auger statistics, Fig.
\ref{fig:Auger}, the point-like nature of the sources is not very 
evident.

A benchmark estimate can help understand the effect: the number of
events detected by an experiment with exposure $\Sigma_{exp}$ (in units
of $\rm km^2~sr~yr$) above some energy threshold $E_{min}$ can be 
written as
$$
N_{ev}(E>E_{th})=\Sigma_{exp} \frac{1}{4\pi}\int_{E_{min}} dE \int_0^{R(E)} 
dr \rho 4\pi r^2~\frac{\Phi(E)}{4\pi r^2} \approx 
$$
\begin{equation}
\approx ~ \Sigma_{exp} \frac{\rho}{4\pi}
\Phi(E_{min}) E_{min} R(E_{min})~, 
\label{eq:bentch}
\end{equation}
where $R(E)$ is the range of cosmic rays with energy $E$ and $\Phi(E)dE$
is the number of particles injected by the source in the energy range
between $E$ and $E+dE$. 

For the AGASA experiment, adopting a threshold $E_{min}=4\times 10^{19}$ 
eV, one has $\Sigma_{exp}=1645~\rm km^2~sr~yr$, $N_{ev}(E>E_{th})=72$ and 
$R_{min}\approx 1000$ Mpc. We have already found that the observed 
doublets and triplets hint at a source density $\sim 10^{-5}\rm Mpc^{-3}$.
Let us denote in general the source density as $\rho = 10^{-5}\rho_{-5}\rm 
Mpc^{-3}$. It is therefore easy to use the previous expression to obtain
$$\Phi(E_{min}) \approx 1.3\times 10^{21} ~ \rm \rho_{-5}~~~eV^{-1}~yr^{-1},$$
consistent with an injection of $\sim 10^{42}\rm erg~s^{-1}$ per source
in the energy range $E>10^{19}$ eV, found with more detailed calculations.

Using now Eq. (\ref{eq:bentch}) one can estimate which exposure is needed 
at energies above $10^{20}$ eV, in order to make sure that we receive at 
least one event per source on average, namely that each source has contributed 
at least one event at the detector. Putting numbers in the equation above, 
and adopting $R(10^{20}~\rm eV)\approx 100~\rm Mpc$, and an injection spectrum 
$E^{-2.6}$, we easily obtain the condition 
\begin{equation}
\Sigma_{exp} > \Sigma_c = 42000~\rho_{-5} ~ \rm km^2~sr~yr.
\end{equation}
The critical exposure $\Sigma_c$ should be compared with 
$\Sigma_{Auger}=35000 ~ \rm km^2~sr~yr$ of Auger after 5 years of operation, 
and with $\Sigma_{EUSO}\sim 10^5 ~ \rm km^2~sr~yr$ after 3 years of operation 
of EUSO. If one interprets our analysis in Fig. \ref{fig:AGASAmulti}
as evidence that the most likely range of source densities should
be $10^{-5}-10^{-4}~\rm Mpc^{-3}$, then EUSO is expected to have 
exposure above the critical. Auger is above the critical exposure
only for the lower limit on the source density, $\rho=10^{-6}~\rm Mpc^{-3}$.

Rephrasing this result, with an experiment with sufficiently large
exposure, at energies above $10^{20}$ eV each source has statistically 
contributed at least one of the detected events. Further increase in 
the experimental exposure at energies above $10^{20}$ eV would not increase 
the number of fainter (more distant) sources, but should rather imply an
increase in the number of events received from each source within the maximum
distance from which the events can reach the Earth. This is a direct 
consequence of the physical meaning of the GZK feature in the cosmic ray 
spectrum, and is of crucial importance for the evaluation of the number 
of sources of UHECRs in the universe. 
In fact, the number of sources can be simply obtained by counting the
multiplets (with multiplicity 1, 2, 3, ...) in the observed data. This 
approach is hardly applicable at lower energies: particles with energy 
$4\times 10^{19}$ eV may reach the Earth from 1 Gpc distance, so that 
the identification of multiplets may in fact be problematic. 

\begin{figure}
\begin{center}
\includegraphics[width=0.9\textwidth]{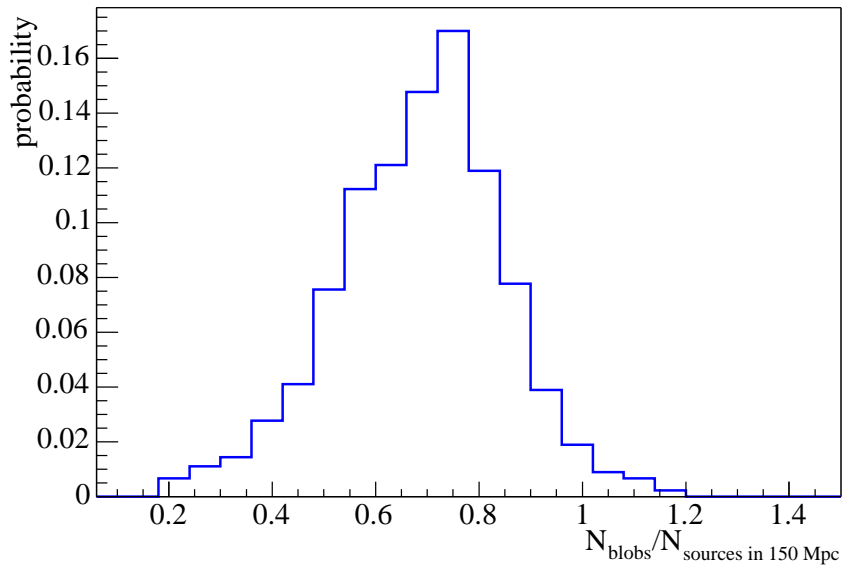}
\caption{Results of the source counting procedure based on the identification
of clusters of events in the sky. On the x-axis the ratio of the number of 
sources counted through the clusters and the {\it real} number of sources
within 150 Mpc is plotted.}
\label{fig:count}
\end{center}
\end{figure}

We implemented a routine for the identification of the clusters of 
events and the counting procedure generates the correct number of 
sources within $L_{95}=150$ Mpc, as illustrated in Fig. \ref{fig:count},
where we plot the ratio of the number of sources obtained from the 
counting procedure and the {\it real} number of sources in the simulation.
Typically $\sim 70-80\%$ of the sources are correctly counted. This 
uncertainty should be compared with the 1-2 orders of magnitude uncertainty 
that can be achieved from current data on the small scale clustering. 
The distance $L_{95}=150~\rm Mpc$ has been evaluated numerically and 
represents 
the distance from which $95\%$ of the events with detected energy above
$10^{20}$ eV start. We stress that the detected energy is affected by 
a statistical error of $\sim 30\%$ that needs to be accounted for, 
when evaluating the distance $L_{95}$.

\section{Direct measurement of a source distance, spectrum and intrinsic 
luminosity} \label{sec:single}

The large statistics of events that will likely be available with future
UHECR observatories will provide us with the unique opportunity to see 
a single source directly and have information about its injection spectrum,
distance and luminosity. As discussed in the previous section, several point
sources are expected to appear in EUSO as multiplets with multiplicity 
$\sim 20-70$ at energies above $10^{20}$ eV. This implies that the spectrum 
of UHECRs from those sources can be reconstructed to some extent. If the 
threshold for triggering of events in EUSO is lowered to $5\times 19^{19}$,
even if with some efficiency factor, this would make the spectral 
reconstruction even more powerful. For the Auger telescope these conditions
are supposed to be less probable, but it may happen that there is a nearby 
powerful source for which the analysis can be carried out.

\begin{figure}
\begin{center}
\includegraphics[width=1.1\textwidth]{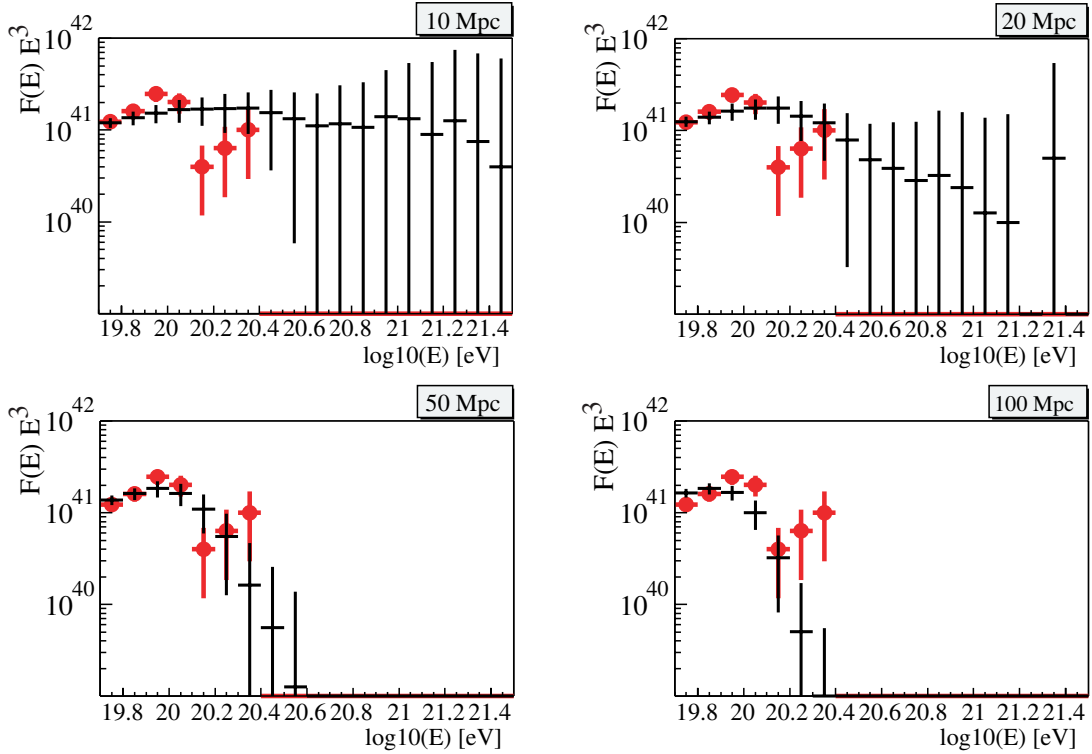}
\caption{Spectrum of a cluster of events (thick dots with error bars)
and the spectra of sources at distances $10,~20,~50~100$ Mpc with the 
same number of events (arbitrary units).}
\label{fig:blobfig}
\end{center}
\end{figure}

It is worth recalling that the GZK feature in the spectrum of single sources
appears to be much more pronounced than in the diffuse flux, since the latter
derives from the superposition of the contributions of the spectra of many 
sources with the GZK feature at different energies. Therefore, studying the
spectra of isolated sources it is possible in principle to infer the injection
spectrum and/or the distance to the source. A practical example is provided
here, in which we use the events that are associated in Fig. 
\ref{fig:EUSO_4panels} with the higher multiplicity cluster of events
in the realization, identified as a source. Note that in this realization, the
cluster only contained 20 events at energies larger than $10^{20}$ eV. 
This is one of the most pessimistic 
situations, since in many realizations clusters containing up to 70 events
are found. In Fig. \ref{fig:blobfig} we 
plot the spectrum of the events above $5\times 10^{19}$ eV coming from the 
cluster (thick dots with error bars) together with four mock (simulated) 
spectra from sources with the same events statistics at four different 
distances ($10,~20,~50,~100~\rm Mpc$). The simulated (mock) spectra from the 
source are averaged over many realizations in order to estimate the error 
bars due to fluctuations in photopion production.
For a source at $10\rm Mpc$ we predict a larger number of events 
at high energies (assuming that the maximum energy at the source is 
large enough). A fit by eye would suggest that the most likely distance
to the source would be $50-100$ Mpc. Moreover, for a total of 120 events
above $5\times 10^{19}$ eV (observed) one would expect that above
$10^{20}$ eV the events would be $35\pm 6$ for a distance of 10 Mpc, 
$32\pm 5$ for a source at 20 Mpc, $21\pm 5$ for a source distance of 50 Mpc
and $10\pm 3$ for a source distance of 100 Mpc, to be compared with 
the observed 22 events in the cluster of events identified as a source,
which in the simulation sits at 48 Mpc. The identification of $\sim 50$ 
Mpc as the likely distance of the source seems to be rather easy to achieve.

It is important to stress that while the current data allow one to speculate
that the lack of association of the Fly's Eye event at $3\times 10^{20}$
eV with any kind of source within 30 Mpc may be due to some unknown
carrier, insensitive to the photopion production energy losses, the 
measurement of the spectrum of the source should pin down the source
distance, independently of its direct identification at other
wavelengths. 

\section{The role of the Intergalactic Magnetic Field}
\label{sec:magnetic}

Much debate exists in current literature about the role of intergalactic
magnetic fields upon the propagation of UHECRs from their sources to the
Earth. This is a debate hard to settle since our knowledge of extragalactic
magnetic field strength and structure is very poor. We have a few pieces
of information which are however difficult to fit in a single puzzle. In
this section we briefly summarize these pieces of information and we make
the attempt to establish the boundaries of the region of applicability of
the results presented in this paper. 

The strongest limits on the intergalactic average magnetic field presently
come from measurements of the Faraday rotation of distant sources. As found
in \cite{burles}, the limits are at the level of $10^{-9}$ Gauss for a 
coherence length of $50$ Mpc, but it is worth to keep in mind that the large 
fluctuations found in \cite{burles} make this limit very uncertain. 

There is at present no evidences that there are magnetic fields in the
intergalactic medium outside large scale structures, such as galaxies 
and clusters of galaxies. In galaxies there are plenty of measurements
of magnetic fields. In particular, in our own galaxy the magnetic field
appears to be in rough equipartition with the thermal and cosmic ray
content. Galactic magnetic fields are however hardly of any relevance
for the propagation of cosmic ray protons with energy above $4\times
10^{19}$ eV: the Larmor radius of these particles is in fact much larger
than the size of a Galaxy and moreover the volume filling factor of 
the universe in the form of galaxies is extremely small, roughly $10^{-5}$.

In clusters of galaxies, there are two separate avenues through
which magnetic fields have been inferred, namely through measurements of
the Faraday rotation of intracluster sources, and from observation of
nonthermal radio and hard X-ray emissions. Radio emission is the result
of synchrotron emission of relativistic electrons in the intracluster
magnetic field, while the hard X-ray emission, observed only from a few
clusters of galaxies, has a more uncertain interpretation. However,
the most straightforward explanation for the X radiation is that it
is the result of Inverse Compton Scattering (ICS) of the same electrons
responsible for the diffuse radio emission against the photons of the
cosmic microwave background. If this is the right interpretation, then
the magnetic field in the intracluster medium of a few clusters can 
be inferred to be in the range $0.1-1\mu G$. Several avenues have been 
proposed to avoid the apparent discrepancy between these values and
those measured through Faraday rotation, which are typically larger
\cite{kronberg}. 
The latter in particular depend upon the electron density along the 
line of sight, relatively well known from observations of the thermal 
X radiation, and the coherence length of the field, which is unknown.

Note that the thermal energy density in a cluster of galaxies is 
$$E_{th}=(3/2) n_{gas} k T \approx 2\times 10^{-11} 
\frac{n_{gas}}{10^{-3}~\rm cm^{-3}}~\frac{T}{10^8~\rm K} \rm erg~cm^{-3},$$
to be compared with the magnetic energy density $E_B=B^2/8\pi=4\times
10^{-14} (B/\mu G)^2 ~\rm erg~cm^{-3}$, about 500 times less than the
thermal energy density, despite the fact that clusters of galaxies
are virialized objects.

Magnetic fields with strength of $\mu G$ over a Mpc scale do 
affect the propagation of UHECRs. However, clusters of galaxies fill
about $10^{-5}$ of the volume of the universe, therefore it is very 
unlikely that they can appreciably modify the spectra of diffuse UHECRs
received at the Earth.

From the discussion above, we conclude that the only magnetic fields that
can affect the propagation of UHECRs are the intergalactic fields outside
galaxies and clusters of galaxies. For the sake of clarity, let us 
split the discussion in two parts, one concerning the magnetic field in
the intergalactic space on large scales, and the other concerning 
our cosmic neighborhood, the so-called local supercluster (LSC). 

On the large scales which are of relevance for the propagation of 
UHECRs with energy below $4\times 10^{19}$ eV, the limits on the 
magnetic field are as described above and depend upon the reversal
scale of the field $L_c$. The results presented here remain valid as 
long as the angular deflection during the propagation along a 
path of length $D$ stays smaller than the angular resolution of
an experiment $\theta_{exp}$. In other words, the following 
condition has to be fulfilled:
$$B < 10^{-10} \left(\frac{E}{4\times 10^{19}eV}\right)
\left(\frac{\theta_{exp}}{2^o}\right)
\left(\frac{D}{1000\rm Mpc}\right)^{-1/2} 
\left(\frac{L_c}{1\rm Mpc}\right)^{-1/2} ~ \rm Gauss.
$$
Clearly the most severe constraints concern the case of AGASA where
in order to collect enough statistics of events one has to consider
particles with energy above $4\times 10^{19}$ eV, whose loss length
is $\sim 1000$ Mpc. In this case, our results are valid if the 
magnetic field is smaller than $1.3\times 10^{-10}$ Gauss, if 
$L_c=1 \rm Mpc$ and $\theta_{exp}=2.5^o$. Whether this is a 
reasonable value for $L_c$ is debatable since it depends crucially
on the model for the magnetic field in the intergalactic medium.
If the field is of cosmological origin, then $L_c=1 \rm Mpc$
may be a reasonable assumption. On the other hand, if the intergalactic 
medium is polluted by the magnetic field spilled out of galaxies through
winds, a more reasonable value would be $L_c\sim 100$ kpc. In 
this case however it is misleading to think of the universe as filled by a
homogeneous magnetic field with some coherence length. It is probably more 
correct to picture it in terms of magnetized bubbles with relatively strong 
magnetic fields surrounded by regions where there is no field \cite{beregaz}. 
In any case, if we take $L_c=100$ kpc, then our analysis remains valid at 
least for $B<4\times 10^{-10}$ Gauss, and this is probably a too strong 
constraint. 
These values are in the same region of magnetic fields constrained by 
Faraday rotation measurements, which leads us to believe that our results
may be considered sufficiently general.

We consider now our cosmic neighborhood and we address the issue of
whether strong fields can be present there. The only evidence for
magnetic fields in superclusters comes from observations of a tenuous
diffuse radio emission from a {\it bridge} connecting the Coma
cluster with Abell 1367, detected at 360 MHz \cite{kim89}. The two
clusters are $\sim 40$ Mpc away from each other, but it is important 
to stress that the so-called bridge extends only for $1.4~h_{75}^{-1}$
Mpc out of the core of the Coma cluster, whose virial radius
is $\sim 3$ Mpc (here $h_{75}$ is the Hubble constant in units of 
$75 \rm Km~s^{-1}~Mpc^{-1}$). The radio emission is a clear indication
of the presence of a magnetic field, but by itself cannot provide
the value of this field, since the same radio flux can be achieved by 
changing both the magnetic field and the electron density.
An assumption which is often made in 
Radio Astronomy is to assume that the magnetic field is in equipartition
with the relativistic electrons and protons in the medium, the ratio
of the two being completely unknown. If one accepts this assumption, 
and if a guess about the ratio of electrons to protons is made (in 
\cite{kim89} this ratio is taken equal to unity) and assuming a 
spectrum for the radiating electrons, the authors derive a magnetic
field of $\sim 0.6\mu G$ . Aside from the many assumptions that 
need to be used to derive this result, it should be kept in mind
that the bridge extends over regions of space which are much smaller
than the distance between Coma and A1367 and in fact smaller than 
the virial radius of Coma. From the point of view of the propagation 
of UHECRs in superclusters this is hardly an evidence for $\mu G$ 
magnetic fields in particular if extended to all superclusters. 

The local supercluster is a $\sim 40$ Mpc long elongated structure 
with the Virgo cluster of galaxies in its center and the local 
group (with our Galaxy in it) at its edge. Its average overdensity
is $\sim 2$, typical of large scale structures that have not reached 
yet the virial equilibrium. Aside from the region around M87, a 
radio galaxy in the center of the Virgo cluster, there is no other
evidence for magnetic fields in the LSC. In the
likely assumption that the supercluster is not a magnetically
dominated structure, namely that the energy density in the field
is smaller than the thermal energy density, the condition 
$B\ll 0.1\mu G$ on the magnetic field must hold. Here we assumed that 
the baryon density is twice that in the background and that the 
temperature in the supercluster is $\sim 10^6$ K. 

The propagation of cosmic rays with energy above $4\times 10^{19}$ eV
would certainly be affected by magnetic fields of the order of 
$10^{-2}-10^{-1}\mu G$ in the LSC. On the other hand one should keep in 
mind an important fact which is often ignored in Montecarlo calculations of
the propagation in strong local fields, namely that the flux of 
CRs at energies below $4\times 10^{19}$ eV can reach the Earth from 
cosmological distances. UHECRs in the energy range 
around $4\times 10^{19}$ eV have a loss length comparable with the 
size of the universe, say 1 Gpc. If $j_{CR} (E)$ is the cosmic ray 
emissivity in the intergalactic space, the corresponding emissivity 
in the LSC will be in general amplified by some
factor $\xi>1$. The flux of UHECRs contributed by the LSC
is therefore $\Phi_{LSC}\sim \xi~j_{CR}~R_{SLC}$ if the field is not
large enough to fall in the diffusive regime of cosmic ray propagation. 
The flux contributed by the rest of the universe is however
$\sim j_{CR}~L_{loss}$ (with $L_{loss}$ the loss length), so that the 
ratio of the two is $\sim 10^{-2}~\xi$ at energies $\sim 4\times 10^{19}$ eV.
As stressed above, the local overdensity in the LSC is $\sim 2$, therefore
we expect $\xi$ to be of the same order. It follows that the flux of UHECRs
at the Earth at energies $\sim 4\times 10^{19}$ eV is dominated by 
the universe outside the LSC. It is therefore necessary to simulate
the propagation of UHECRs over the whole universe for the purpose of 
understanding spectrum and anisotropies of cosmic rays around
$\sim 4\times 10^{19}$ eV. The numerical investigations in \cite{isola}
concerning the small scale anisotropies from sources in the LSC and
in \cite{ensslin}, that make an attempt to adopt a realistic local 
structure of the magnetic field and sources, do not include this 
important effect.

We summarize the discussion above as follows:

1) the magnetic field in the universe is likely to have a bubble-like
structure, with regions in which the field is relatively intense, surrounded
by large volumes where the magnetic pollution is not present. If 
the magnetic field is of cosmological origin, the region of values of the
average magnetic field for which the analysis presented here can be applied 
overlaps with the upper limits obtained from Faraday rotation measures.
If the correlation, claimed in \cite{tkachev1}, between UHECRs (with 
energy between $2\times 10^{19}$ eV and $10^{20}$ eV) and distant BL Lac
objects is correct, then a clear evidence of very weak magnetic fields over 
cosmological scales follows \cite{beregaz}. 

2) A magnetic field as strong as $\sim 10^{-8} \rm Gauss$ (equal to the 
equipartition field) in the LSC would
still be consistent with upper limits derived from measurements of Faraday 
rotation. However at present there is no indication or evidence that 
such a magnetic field is in fact there. Moreover, as found in 
\cite{sato1,sato2}, such a field may be hard to reconcile with the 
multiplets observed with AGASA.
It is therefore important to understand which information can be inferred 
from small scale anisotropies of UHECRs in the simplest possible scenario,
namely in the regime of weak fields.
We will consider some implications of a possible weak local magnetic field 
in a forthcoming paper, since as is well known, cosmic rays remain a 
precious tool to investigate the presence of magnetic fields in the universe.

\section{Discussion and Conclusions}
\label{sec:conclude}

We investigated the potential of present and future experiments to 
identify the sources of UHECRs, by using the combined information 
on the spectrum and small scale anisotropies in the arrival directions. 

While the spectrum of UHECRs can be used to obtain the total rate of
energy injection per unit volume, corresponding to $\sim 6\times 10^{44} 
\rm erg~yr^{-1}~Mpc^{-3}$ above $10^{19}$ eV, the existence of small
scale anisotropies in the AGASA data allows us to infer the approximate 
density of sources, with an approximation of less than two orders of magnitude 
around the average value of $\sim 10^{-5}~\rm Mpc^{-3}$. The luminosity per
source is therefore  $L_{source}\approx 2\times 10^{42}\rm erg~s^{-1}$
above $10^{19}$ eV, for a best fit injection spectrum $E^{-2.6}$. 
If extrapolated to lower energies, the corresponding luminosity 
increases to $L_{source}\approx 2\times 10^{48}\rm erg~s^{-1}$ above
1 GeV. This number suggests that the sources should in fact be very 
bright, unless a mechanism is found to limit the acceleration 
only to very high energy particles: for instance, for the case
of acceleration at relativistic shocks moving with Lorentz factor 
$\Gamma_{sh}$, a minimum energy of $\Gamma_{sh}^2 m_p c^2$ is expected, 
which may decrease the required luminosity appreciably. 

In case of evolution of the sources with a functional dependence
$(1+z)^4$, an injection spectrum $E^{-2.4}$ can fit the data, at least 
at energies above $10^{19}$ eV, requiring a luminosity per source 
above $10^{19}$ eV which is basically the same as in the non evolutionary 
case (but two orders of magnitude smaller if extrapolated down to 
$1$ GeV). The small scale anisotropies are not appreciably affected by 
the evolution of the sources, because they are mainly determined by nearby 
sources ($z\leq 0.3$), and their number is fixed by the highest energy 
events. In this case the density quoted above should be interpreted as 
the local density of sources. 

The upcoming experiments such as the Pierre Auger Observatory and the 
EUSO Observatory will open many new avenues to determine the nature of 
the sources of UHECRs. We discussed here the potentials of these two
experiments, with the help of numerical simulations for the propagation 
of UHECRs, that allowed us to simulate the expected statistics of events.

The two point correlation function is used to study the clustering 
properties of the events generated from sources having the density obtained
from the analysis of the AGASA data. We find that the correlation function with
the statistics of events expected with Auger shows a very pronounced peak 
at small scales, confirming the point-like nature of the sources. 

For the case of Auger, the two point correlation function of events
with energy above $4\times 10^{19}$ may not be the best 
tool to infer the point-like nature of the sources. In fact in the
$\sim 2$ degrees of average angular resolution of the experiment
doublets and triplets of events occur at random, not necessarily due 
to the point-like nature of the sources, but rather due to the abundance of 
events. In other words the multiplets often result from the contribution 
of different sources. 

A better result is obtained by calculating the two point correlation
function limited to the events with energy above $10^{20}$ eV. In this 
case the calculated error bars appear to be small enough to allow
the source counting with an uncertainty of one order of magnitude 
or less. A corresponding uncertainty has to be expected in the
estimate of the luminosity of single sources. 

The analysis based upon the calculation of the two point correlation function
was also carried out for the EUSO expected statistics of events above 
$10^{20}$ eV. Besides the peak at small angular scales, the two point 
correlation function shows now a series of peaks that are not due to 
fluctuations in the number of events as they would result from propagation. 
These peaks are due to Poisson noise around the average of the correlation 
function at large angles,
due to fluctuations in the number of sources in a given area. The wiggles 
disappear if one averages the correlation function over a large number of 
realizations of the source distribution. Although the peaks are washed out 
by statistical averaging, the information about the sources does not 
disappear, but remains in the amplitude of the fluctuations around the mean 
of the correlation function. The peaks provide us with another valuable tool 
to estimate the number density of the sources, with an uncertainty of at most
one order of magnitude. 

The important insight behind the appearance of this structure in the 
two point correlation function is that a {\it critical exposure} exist, which
depends upon the source density in the sky, such that experiments with 
exposures larger than the critical one are able to receive at least 
one event from each source. This is the reason why the peaks introduced
above do not appear in smaller experiments but start to show
up in EUSO if the source density is at the level of $10^{-5}~\rm Mpc^{-3}$. 

This insight allows for a direct counting of the sources in the sky through
the identification of the clusters of events associated with each source.
Our simulations show that this procedure should allow one to count the
sources and therefore determine their density with uncertainties 
within $20-30\%$.

The nearby sources appear as blob-like structures in the sky, and some
of them have very high multiplicity, up to $\sim 50-70$ at energies
above $10^{20}$ eV. It is therefore proposed that the spectrum of 
the sources can be measured for the first time, also allowing the
evaluation of the source distance. If this goal can be achieved, 
it would represent the first clear identification of the source 
itself, independently of the detection of the same source at other
wavelengths. 

The value of the source density that has been found here is comparable
with the typical density of cosmic objects such as rich clusters of galaxies
and active galactic nucleei. While the former cannot accelerate particles
to the highest energies, active galaxies of some type might do it. 
It is therefore worth pursuing the investigation of the acceleration 
mechanisms in these sources.

\section*{Acknowledgments} We thank Roberto Aloisio, Venya Berezinsky, 
Stefano Gabici, Angela Olinto and Mario Vietri for many useful discussions 
and ongoing collaboration. We are also grateful to William Burgett for a 
useful correspondence, to Alessandro Petrolini for valuable comments to
the manuscript and to Andrea Ferrara and Evan Scannapieco for a very useful 
discussion on correlation functions. We are also grateful to the anonymous 
referee for constructive comments. This work was partially supported 
through grant COFIN-2002 at Arcetri.

%\clearpage


\begin{thebibliography}{99}

\bibitem{tkachev1}
P.G. Tinyakov and I.I. Tkachev, JETP Lett. {\bf 74} (2001) 445; 
Pisma Zh. Eksp. Teor. Fiz. {\bf 74} (2001) 499

\bibitem{tkachev2}
D. Gorbunov, P. Tinyakov, I. Tkachev and S. Troitsky, Astrophys. J. Lett.
{\bf 577} (2002) L93

\bibitem{daniel1}
D. De Marco, P. Blasi and A.V. Olinto, Astropart. Phys., in press 
(preprint astro-ph/0301497)

\bibitem{HIRES1}
T. Abu-Zayyad, et al., preprint astro-ph/0208301

\bibitem{HIRES2}
T. Abu-Zayyad, et al., preprint astro-ph/0208243

\bibitem{uchihori}
Y. Uchihori, M. Nagano, M. Takeda, M. Teshima, J. Lloyd-Evans and 
A.A. Watson, Astropart. Phys. {\bf 13} (2000) 151

\bibitem{harari}
D. Harary, S. Mollerach and E. Roulet, JHEP {\bf 2} (2000) 035

\bibitem{tom}
H. Goldgerg and T.J. Weiler, Phys. rev. {\bf D64} (2001) 056008

\bibitem{will}
W.S. Burgett and M.R. O'Malley, Phys. Rev. {\bf D67} (2003) 092002

\bibitem{dubovsky}
S.L. Dubovsky, P.G. Tinyakov and I.I. Tkachev, Phys. Rev. Lett. {\bf 85}
(2000) 1154

\bibitem{fodor}
Z. Fodor and S.D. Katz, Phys.Rev. {\bf D63} (2001) 023002

\bibitem{auger}
J. W. Cronin, Proceedings of ICRC 2001 (2001).

\bibitem{EUSO} see http://www.euso-mission.org

\bibitem{sato1}
H. Yoshiguchi, S. Nagataki, S. Tsubaki and K. Sato, Astrophys. J. {\bf 586}
(2003) 1211

\bibitem{sato2}
H. Yoshiguchi, S. Nagataki and K. Sato, preprint astro-ph/0302508

\bibitem{Stanevmag} 
T. Stanev, preprint astro-ph/0108338

\bibitem{isola}
C. Isola and G. Sigl, Phys. Rev. {\bf D66} (2002) 3002

\bibitem{ensslin}
G. Sigl, F. Miniati and T. Ensslin, preprint astro-ph/0302388 

\bibitem{Blanton} M. Blanton, P. Blasi and A.~V. Olinto, Astropart. 
Phys. {\bf 15} (2001) 275

\bibitem{Blumenthal:nn} 
G.~R.~Blumenthal, Phys. Rev. {\bf D1} (1970) 1596.

\bibitem{czs} 
M.~J.~Chodorowski, A.~A.~Zdziarski, and M.~Sikora, Astrophys. J.
{\bf 400} (1992) 181.

\bibitem{BS00}
P. Bhattacharjee and G. Sigl, Phys. Rept.  327 (2000) 109

\bibitem{Stanev}
T.~Stanev, R.~Engel, A.~Mucke, R.~J.~Protheroe and J.~P.~Rachen,
Phys.\ Rev.\ D {\bf 62} (2000) 093005.

\bibitem{beregrig}
V. Berezinsky and S. Grigorieva, Astron. Astroph.  199 (1988) 1

\bibitem{berebook} V. S. Berezinsky, S.V. Bulanov, V. A. 
Dogiel, V. L. Ginzburg,  and V. S. Ptuskin,  Astrophysics of Cosmic 
Rays, (Amsterdam: North Holland, 1990)

\bibitem{bereAGN}
V. Berezinsky, A. Z. Gazizov and S. Grigorieva, preprint astro-ph/0210095

\bibitem{bgg2002}
V. Berezinsky, A. Z. Gazizov and S. Grigorieva, preprint hep-ph/0204357

\bibitem{takeda}
M. Takeda, et al., Astrophys. J. {\bf 522} (1999) 225

\bibitem{agasa1}
N. Hayashida et al., Astrophys. J. {\bf 522} (1999) 225

\bibitem{sommers}
P. Sommers, Astropart. Phys. {\bf 14} (2001) 271

\bibitem{burles}
P. Blasi, S. Burles and A.V. Olinto, Astrophys. J. Lett. {\bf 514} (1999) 79

\bibitem{kim89}
K.-T. Kim, P.P. Kronberg, G. Giovannini and T. Venturi, Nature {\bf 341}
(1989) 720

\bibitem{beregaz}
V. Berezinsky, A. Z. Gazizov and S. Grigorieva, preprint astro-ph/0302483 

\bibitem{kronberg}
T.E. Clarke, P.P. Kronberg and H. B\"{o}hringer, Astrophys. J. Lett.
547 (2001) 111

\end{thebibliography}
\end{document}